\documentclass{ws-procs975x65}

\def \la{\mathrel{\mathchoice   {\vcenter{\offinterlineskip\halign{\hfil
$\displaystyle##$\hfil\cr<\cr\sim\cr}}}
{\vcenter{\offinterlineskip\halign{\hfil$\textstyle##$\hfil\cr
<\cr\sim\cr}}}
{\vcenter{\offinterlineskip\halign{\hfil$\scriptstyle##$\hfil\cr
<\cr\sim\cr}}}
{\vcenter{\offinterlineskip\halign{\hfil$\scriptscriptstyle##$\hfil\cr
<\cr\sim\cr}}}}}

\begin{document}

\title{AMMONIA AS A TRACER OF FUNDAMENTAL CONSTANTS}

\author{CHRISTIAN HENKEL$^*$ and KARL M. MENTEN}

\address{Max-Planck-Institut f{\"u}r Radioastronomie\\ 
Auf dem H{\"u}gel 69, 53121 Bonn, Germany\\ 
E-mail: chenkel@mpifr-bonn.mpg.de, kmenten@mpifr-bonn.mpg.de}

\author{MICHAEL T. MURPHY}

\address{Center for Astrophysics and Supercomputing, Swinburne University of Technology\\
Mail H39, PO Box 218, Victoria 3122, Australia\\
E-mail: murphy@swin.edu.au}

\author{VICTOR V. FLAMBAUM}

\address{School of Physics, Univ. of New South Wales, Sydney, N.S.W. 2052, Australia\\
E-mail: flambaum@newt.phys.unsw.edu.au}

\author{SERGEI A. LEVSHAKOV}

\address{Ioffe Physical Technical Institute, Polytekhnicheskaya Str. 26, 194021 St. Petersburg, Russia\\
lev@astro.ioffe.rssi.ru}

\author{ALEXANDER V. LAPINOV}

\address{Institute for Applied Physics, Uljanov Str. 46, 603950 Nizhny Novgorod, Russia\\
E-mail: lapinov@appl.sci-nnov.ru}

\author{PAOLO MOLARO}

\address{INAF-Osservatorio Astronomico di Trieste, Via G.B. Tiepolo 11, 34131 Trieste, Italy\\
E-mail: molaro@oats.inaf.it}

\author{JAMES A. BRAATZ}

\address{National Radio Astronomy Observatory, 520 Edgemont Road, Charlottesville, VA 22903, 
USA \\ E-mail: jbraatz@nrao.edu}

\begin{abstract}
Observing inversion lines of ammonia (NH$_3$), complemented by rotational lines of NH$_3$ 
and other molecular species, provides stringent constraints on potential variations
of the proton-to-electron mass ratio, $\mu$. While a limit of $|\Delta$$\mu|$/$\mu$ $\sim$ 
10$^{-6}$ is derived for a lookback time of 7$\times$10$^9$\,yr, nearby dark clouds
might show a significant variation of order (2--3)$\times$10$^{-8}$, possibly being
related to chameleon fields. The detection of radio-loud quasars with strong 
molecular absorption lines at redshifts $z$ $>$ 1 as well as the identification of a 
larger sample of nearby molecular clouds with exceptionally narrow lines ($\Delta V$  
$<$ 0.2\,km\,s$^{-1}$) would be essential to improve present limits and to put the 
acquired results onto a firmer statistical basis.
\end{abstract}


\bodymatter

\section{Introduction}\label{sec1}
Comparing redshifts of various spectral lines, observed toward the same distant 
highly redshifted object, has the potential to yield important contraints on temporal 
variations of fundamental constants of the standard model over timescales of billions 
of years, even surpassing the age of the solar system. A pre-condition is that the lines 
in question have different dependencies on the respective constant. Among the most commonly 
studied fundamental parameters are the dimensionless fine structure constant, $\alpha$, and the 
proton-to-electron mass ratio, $\mu$ (e.g., Uzan 2003; Garc{\'i}a-Berro et al. 2007; Dent 2008). 
It may be possible that variations in $\mu$ are more pronounced than those in $\alpha$ 
(e.g., Flambaum 2008). Therefore, studies contraining $\mu$ may be particularly
rewarding.

\section{The role of ammonia}\label{sec2}

Making use of the different dependencies of electronic, vibrational, and rotational
line frequencies of molecules on the proton-to-electron mass ratio $\mu$, limits on variations
are commonly derived from H$_2$ spectra of damped Lyman $\alpha$ systems measured
in absorption toward quasars. To date, these observations indicate that $\mu$ is unchanged
to an accuracy of order 10$^{-5}$ over the last 80\% of the age of the Universe (e.g.,
Malec et al. 2010). 

Ammonia (NH$_3$), having being detected for the first time at significant redshifts by 
Henkel et al. (2005, 2008), opens up a new avenue to constrain $\mu$ over large time scales. 
Its inversion transitions are strongly dependent on $\mu$ (J. N. Chengalur, priv. comm.; 
Flambaum \& Kozlov 2007), with a fractional change in frequency $\Delta \nu_{\rm 
inv}$/$\nu_{\rm inv}$ corresponding to \hbox{--4.46}$\Delta \mu$/$\mu$. For rotational 
lines, $\Delta \nu_{\rm rot}$/$\nu_{\rm rot}$ = --$\Delta \mu$/$\mu$. Therefore, 
comparing NH$_3$ inversion line redshifts with redshifts of rotational transitions 
should provide sensitive limits on the variation of $\mu$.

\section{B0218+357 and PKS\,1830--211}\label{sec3}

The two redshifted sources, toward which ammonia has been detected in absorption, are the main 
gravitational lenses of the quasars B0218+357 and PKS\,1830-211. Flambaum \& Kozlov (2007) 
combine the three detected NH$_3$ absorption spectra from B0218+357 with rotational spectra 
of CO, HCO$^+$, and HCN to place a limit of $|\Delta \mu|$/$\mu$ = (0.6 $\pm$ 1.9) $\times$
10$^{-6}$ for a lookback time of 6 $\times$ 10$^9$\,yr (redshift $z$ = 0.68). Accounting in 
detail for the velocity structure of the line profiles, Murphy et al. (2008) reanalyzed
the ammonia data in combination with newly obtained high-signal-to-noise rotational
spectra of HCO$^+$ and HCN. This yields $|\Delta \mu|$/$\mu$ $<$ 1.8 $\times$ 10$^{-6}$ at 
a 95\% confidence level.

While B0218+357 provided first useful limits, PKS\,1830-211 is by far more suitable when
trying to constrain $\mu$. With ten (instead of three) detected NH$_3$ inversion lines and 
a forest of rotational transitions at nearby frequencies, a higher accuracy can be achieved.
From a comparison of the ammonia inversion lines with the NH$_3$ ($J,K$) = (1,0) $\leftarrow$ 
(0,0) rotational transition, Menten et al. (2008) find consistency within 3.8 $\times$ 10$^{-6}$
at a 95\% confidence level. The strength of this study is its focus on lines arising entirely
from one molecular species. However, the frequencies of the inversion lines are much lower
than that of the rotational line by a factor of $\sim$25, which {\it might} cause differences 
in the absorbed background radio continuum. 

Analyzing the ten NH$_3$ inversion lines and a similar number of rotational transitions
from other molecules, Henkel et al. (2009) obtain a 3$\sigma$ limit of $\Delta \mu$/$\mu$ 
= 1.2 $\times$ 10$^{-6}$ for a lookback time of 7 $\times$ 10$^9$\,yr ($z$=0.89). This study 
is based exclusively on optically thin absorption features located within a limited frequency 
band that were observed within a limited time interval, thus minimizing effects caused by a time 
variable and frequency dependent continuum background morphology. Also, no frequency shift 
as a function of excitation is found. Nevertheless, a detailed velocity component analysis 
like that presented by Murphy et al. (2008) for the much smaller data set from B0218+357 has
not yet been performed. 

\section{Nearby dark clouds}\label{sec4}

We can probe the values of fundamental constants at a considerably more accurate level
in quiescent dark clouds of the Milky Way, where line widths $\Delta V$ $\ll$ 1\,km\,s$^{-1}$
are encountered. $\mu$, measured in the drastically different environments of high terrestrial
and low interstellar densities of baryonic matter, is supposed to vary in Chameleon-like
scalar field models, which predict a strong dependence of masses and coupling constants on
the ambient density (Olive \& Pospelov 2008; Upadhye et al. 2010). High spectral resolution 
data of NH$_3$, HC$_3$N, and N$_2$H$^+$ provide a variation of $\Delta \mu$/$\mu$ = (--2.2 
$\pm$ 0.4$_{\rm stat}$ $\pm$0.3$_{\rm sys}$) $\times$ 10$^{-8}$ (Levshakov et al. 2010).

\section{Conclusions}\label{sec5}

Measurements of ro-vibrational H$_2$ quasar absorption spectra yield $|\Delta \mu|$/$\mu$ $<$
10$^{-5}$ over the last 80\% of the age of the Universe. Radio data including NH$_3$ inversion
lines result in $|\Delta \mu|$/$\mu$ $\la$ 10$^{-6}$ over the last 50\% of the age of the Universe.
Similar data obtained from local dark clouds suggest a potential variation of order 
(2--3) $\times$ 10$^{-8}$.

\bibliographystyle{ws-procs975x65}
\bibliography{ws-pro-sample}

\end{document}